\documentclass[aps,amsfonts,amsmath,superscriptaddress,amssymb,prb,twocolumn]{revtex4}

\usepackage{graphicx}
\usepackage{dcolumn}
\usepackage{bm}
\usepackage{longtable}

\begin{document}

\title{INTERFACE FERROMAGNETISM IN A $\rm
SrMnO_{3}/LaMnO_{3}$ SUPERLATTICE}

\author{S. Smadici}
\affiliation{Frederick Seitz Materials Research Laboratory, University of Illinois, Urbana, Illinois 61801}%

\author{B.B. Nelson-Cheeseman}
\affiliation{Materials Science Division, Argonne National Laboratory, Argonne, Illinois 60439}%

\author{A. Bhattacharya}
\affiliation{Materials Science Division, Argonne National Laboratory, Argonne, Illinois 60439}%
\affiliation{Center for Nanoscale Materials, Argonne National Laboratory, Argonne, Illinois 60439}%

\author{P. Abbamonte}
\affiliation{Frederick Seitz Materials Research Laboratory, University of Illinois, Urbana, Illinois 61801}%

\begin{abstract}
Resonant soft x-ray absorption measurements at the O $K$ edge on a
$\rm SrMnO_{3}/LaMnO_{3}$ superlattice show a shoulder at the energy
of doped holes, which corresponds to the main peak of resonant
scattering from the modulation in the doped hole density. Scattering
line shape at the Mn $L_{3,2}$ edges has a strong variation below
the ferromagnetic transition temperature. This variation has a
period equal to half the superlattice superperiod and follows the
development of the ferromagnetic moment, pointing to a ferromagnetic
phase developing at the interfaces. It occurs at the resonant
energies for $\rm Mn^{3+}$ and $\rm Mn^{4+}$ valences. A model for
these observations is presented, which includes a double-exchange
two-site orbital and the variation with temperature of the hopping
frequency $t_{ij}$ between the two sites.
\end{abstract}

\maketitle

\section{Introduction}

Doped $\rm La_{1-x}Sr_{x}MnO_{3}$ (LSMO) has multiple FM, AFM and
canted magnetic orders as a function of doping and
temperature~\cite{1955Wollan,1955Goodenough,1960deGennes} from
superexchange~\cite{1950Anderson} and
double-exchange~\cite{1951Zener,1955Anderson} interactions, which
favor an antiferromagnetic (AFM) insulating phase and a
ferromagnetic (FM) metallic phase, respectively. The FM phase at low
temperatures is near the $x=0.33$ doping, at which the closely
related manganite $\rm La_{1-x}Ca_{x}MnO_{3}$ shows very large
(``colossal") magnetoresistance (CMR)~\cite{1994Jin}.

The wave vectors of AFM and orbital orders in bulk manganites can be
accessed with soft x-ray scattering at the Mn $L_{3,2}$ edges. For
instance, studies were made for $\rm Pr_{1-x}Ca_{x}MnO_{3}$
(Ref.~\onlinecite{2004Thomas}), $\rm La_{2-x}Sr_{x}MnO_{4}$
(Refs.~\onlinecite{2003Wilkins-b,2004Dhesi,2005Wilkins}) and $\rm
La_{2-2x}Sr_{1+2x}Mn_{2}O_{7}$ (Ref.~\onlinecite{2006Wilkins})
manganites. More recent measurements showed that magnetic and
orbital scattering amplitudes are similar~\cite{2009Staub}, studied
the doping dependence~\cite{2010GF}, confirmed the separation in
energy of Ref.~\onlinecite{2009Staub} between the magnetic and
orbital scattering resonances,~\cite{2011Zhou} and studied their
evolution after photoexcitation~\cite{2011Ehrke}. The broader
features of the measurements are obtained in calculations of line
shapes at the Mn $L_{3,2}$ edges with atomic multiplet models of
magnetic~\cite{2005Stojic} and orbital~\cite{2000Castleton}
scattering, and more recently, with a finite-difference
method~\cite{2010OBthesis}. However, the investigation of the bulk
FM phase near the $x=0.33$ doping is not possible with soft x-ray
scattering at the Mn $L$ edges, due to lack of contrast for this
order.

The FM phase can be studied with soft x-ray scattering in $\rm{
(SMO)}_{\emph{n}}/\rm{(LMO)}_{\emph{2n}}$ superlattices (SL), in
which the Sr and La atoms are ordered in $\rm SrMnO_{3}$ (SMO) and
$\rm LaMnO_{3}$ (LMO) layers. External magnetic fields are not
necessary. The SL growth sequence can be used to define the period
and symmetry of a reflection along the $c$-axis. This was
demonstrated for a $n=4$ SL, in which the scattering wave vector was
decreased to a range accessible at the O $K$ edge resonance and a
higher symmetry of a reflection used to probe interface
scattering.~\cite{2007S}

In this work, we have applied these ideas to SL reflections at the O
$K$ and Mn $L_{3,2}$ edges and studied the development of the FM
moment in a shorter superperiod $n=2$ SL with soft x-ray absorption
and scattering. Measurements at the O $K$ edge showed modulated hole
doping at oxygen sites. We have observed scattering at the Mn $L$
edges from the SL interfaces at SL reflection $L=2$, following the
temperature dependence of the FM moment. The symmetry of the SL
reflection allowed us to probe all Mn valences in the interface
layers. In addition to $\rm Mn^{3+}$ valence resonances of bulk AFM
order
measurements~\cite{2004Thomas,2003Wilkins-b,2004Dhesi,2005Wilkins,2006Wilkins,2009Staub,2010GF,2011Zhou,2011Ehrke},
a peak in the resonant line shape, which has not been observed
before, is aligned with the fluorescence yield edge for the $\rm
Mn^{4+}$ valence. We present a model of the x-ray scattering from
the SL interfaces, which includes the temperature dependence of the
double-exchange hopping frequency $t_{ij}$ and the change in the
configurations of the Mn ions in the FM state.

\section{Experiments}

\subsection{Structure}

The $n=2$ SL was grown on the (001) surface of $\rm SrTiO_{3}$ (STO)
at Argonne National Laboratory at $\rm 700~^{\circ}C$ in a $\rm
2\times 10^{-6}~Torr$ ozone pressure, followed by cooling to $\rm
100~^{\circ}C$ and pump down. The structure was $\{r\times \rm
[2(SMO)+4(LMO)]+SMO\}$ with $r=13$ [Fig. 1(a)]. SMO ($a_{\rm
SMO}=3.805 \rm~\AA$, Ref.~\onlinecite{2001Chmaissem}) and LMO
($a_{\rm LMO}=3.99\rm ~\AA$,
Ref.~\onlinecite{1998Rodriguez-Carvajal}) layers on the STO
substrate ($a_{s}=3.905\rm ~\AA$) are under $+2.6~\%$ tensile and
$-2.2~\%$ compressive strain. The surface RMS roughness, measured
with an atomic force microscope, was $\sigma_{s}=\rm 2.85~\AA$. From
hard and soft x-ray reflectivity measurements, the SL superperiod
was $c_{SL}=22.5\pm0.5~\rm \AA$ and the average $c$-axis parameter
for 1 ML (a coverage of one formula unit of SMO or LMO over a
$a_{s}\times b_{s}$ area) was $\rm 3.86\pm 0.05~\AA$.

\begin{figure}
\centering\rotatebox{0}{\includegraphics[scale=0.5]{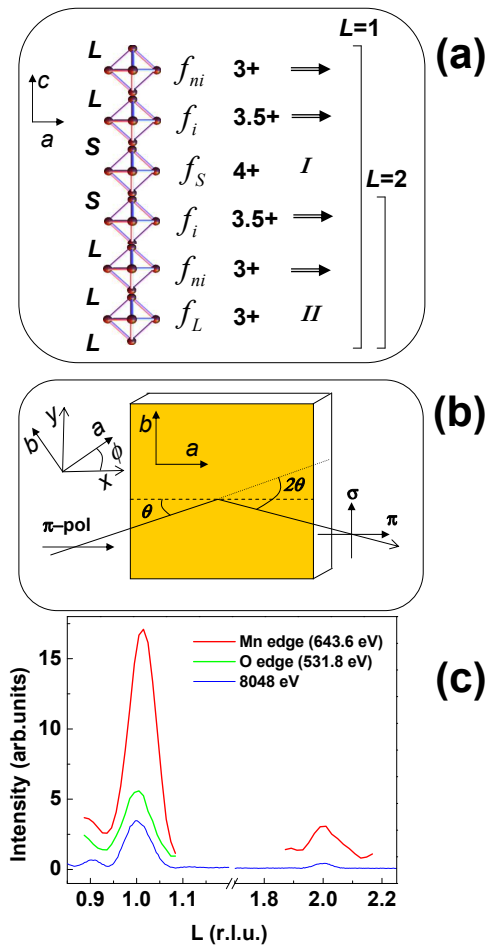}}
\caption{\label{fig:Figure1} (Color online) (a) Sketch of a 6 ML
superperiod of the $\rm (SMO)_{2}/(LMO)_{4}$ superlattice with LaO
($L$) and SrO ($S$) planes, layer form factors $f$, Mn valences in
the $\rm MnO_{2}$ planes estimated from the neighboring $L$ or $S$
planes, a magnetic order in the FM state and periods of the $L=1$
and $L=2$ SL reflections. (b) Scattering geometry for azimuth
$\phi=0^{\circ}$. The scattering vector is along the $c$-axis,
normal to the surface. (c) SL momentum scans at the Mn $L_{3}$, O
$K$ edges and with hard x-rays (8048 eV).}
\end{figure}

\subsection{X-ray absorption}

X-ray absorption spectroscopy (XAS) measurements in fluorescence
(FY) and electron yield (EY) modes were made at undulator beamline
X1B at the National Synchrotron Light Source. The incident light was
$\pi$-polarized and the incidence and detector angles were
$\theta=80^{\circ}$ and $2\theta=110^{\circ}$ [Fig. 1(b)]. The
calculated energy resolution was $\rm 0.39~eV$ and $\rm 0.59~eV$ at
O $K$ (520 eV) and Mn $L_{3}$ (640 eV) edges, respectively.

SL FY and EY measurements at the O $K$ edge show doped holes on the
oxygen sites (Fig. 2). Because the probing depth exceeds the total
SL thickness, FY has contributions from both the SL and the
substrate. In contrast, because of the short electron escape depth,
EY measurements are from the SL top layers only. The shoulder in FY
measurements at 530.3 eV is aligned with the first peak in EY and is
not present in FY measurements of the bare STO substrate. This
shoulder corresponds to doped holes in LSMO
(Ref.~\onlinecite{1997Ju}) and to the $L=1$ scattering peak at 529.6
eV. The peak at 531.8 eV is from the STO substrate. The SL FY and EY
measurements at the O $K$ edge show no discernible variation with
temperature between 300 K and 255 K.

SL FY and EY measurements at the Mn $L_{3,2}$ edges are compared to
FY measurements on bulk samples with different Mn valences
(Refs.~\onlinecite{1991Cramer,2004Morales,2009Lee}) in Fig. 3. SL FY
does not have the sharp peak at the lower energies characteristic of
the $\rm Mn^{2+}$ valence, which shows that the SL has only the $\rm
Mn^{3+}$ and $\rm Mn^{4+}$ valences. The SL FY measurement was
aligned in energy with the average of bulk FY for the $\rm Mn^{3+}$
and $\rm Mn^{4+}$ valences, according to the number of SMO ($\rm
Mn^{4+}$ valence) and LMO ($\rm Mn^{3+}$ valence) layers in one
superperiod. No discernible variation was observed in FY or EY
between 300 K and 255 K.

XAS measurements probe the average valence of O and Mn atoms in the
SL. To discern a variation with temperature in different SL layers
it is necessary to turn to scattering.

\subsection{Resonant soft x-ray scattering}

Resonant soft x-ray scattering (RSXS) measurements were made at the
same beamline with an ultra-high vacuum diffractometer. For other
RSXS experiments on bulk and SL at this endstation see
Refs.~\onlinecite{2005Abbamonte,2007S,2009S,2011Lee,2011S,2012S}.
Detection was for both $\pi-\pi$ and $\pi-\sigma$ channels. The
scattering momentum in the reflectivity geometry $Q=(0,0,2\pi
L/c_{SL})$ is indexed with respect to the SL superperiod $c_{SL}$.
The energy resolution was $\rm 0.20~eV$ and $\rm 0.34~eV$ at O $K$
(520 eV) and Mn $L_{3}$ (640 eV) edges, respectively. The sample was
cooled in zero magnetic field and scattering measurements for $L=1$
at the O $K$ edge and for $L=1,2$ at the Mn $L_{3,2}$ edges [Fig.
1(c)] were made at different temperatures (Figs. 2 and 4).

FM order in metallic films has been studied with x-ray resonant
magnetic scattering of linearly~\cite{1990Kao} and
circularly~\cite{1994Kao} polarized light at the Fe and Co $L$
edges. The FM order in a Ag/Ni SL has been investigated with
circularly polarized light~\cite{1995Tonnerre}. However, unlike
previous studies, where an external magnetic field was applied to
separate the magnetic and charge scattering, the SL FM order is
accessed here at SL reflections with no applied magnetic fields.

\begin{figure}
\centering\rotatebox{0}{\includegraphics[scale=0.5]{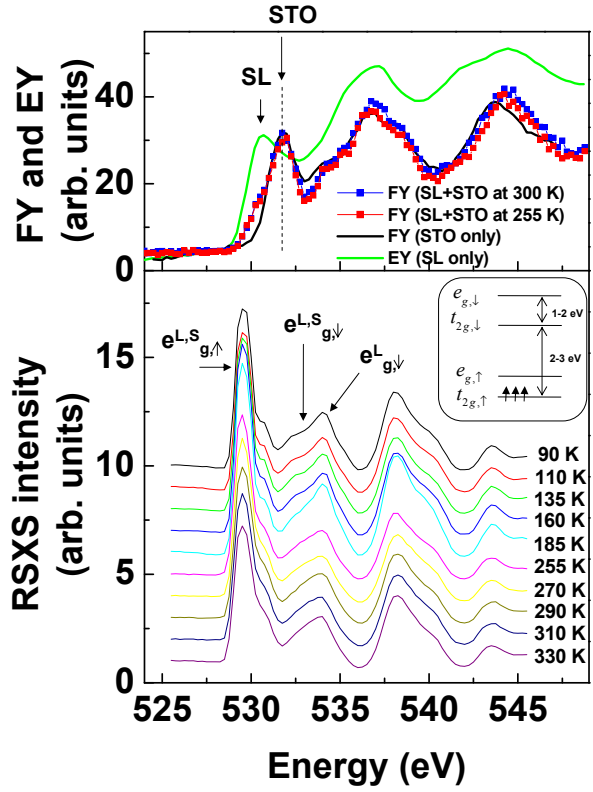}}
\caption{\label{fig:Figure1} (Color online) Top: FY at the O $K$
edge for the SL on STO at different temperatures and the bare STO
substrate (black line), compared to EY measurements. FY measurements
were aligned with the linear transformation
$FY_{plotted}=aFY_{measured}+b$, where $a$ and $b$ are constants.
Bottom: Temperature dependence of resonant scattering at $L=1$. The
scans have been normalized to the pre-edge values and shifted
vertically for clarity. The inset shows the order of the energy
levels for SMO in the ground state, with the Jahn-Teller splitting
neglected. The upper level $e_{g,\downarrow}$ is split in LMO.}
\end{figure}

\subsubsection{O $K$ edge line shape}

The spatial modulation in the density of holes doped on the oxygen
sites can be observed with RSXS. Specifically, the O $K$ edge line
shape for $L=1$ scattering shows a peak close to the energy of the
shoulder in FY measurements (Fig. 2).

Since the order of levels in the RSXS line shape at the O $K$ edge
follows that of the ground state, the RSXS line shape can be
analyzed using the hybridization between O $p$ and Mn $e_{g}$ levels
in the ground state. XAS experiments and calculations at the Mn $L$
edge (Ref.~\onlinecite{1992Abbate}) give a crystal field splitting
$10Dq$ between the $e_{g}$ and $t_{2g}$ levels of 1.5 eV for bulk
LMO and 2.4 eV for SMO. Fig. 2 (inset) shows the unoccupied
$e_{g,\uparrow}$, $t_{2g,\downarrow}$ and $e_{g,\downarrow}$ levels
in SMO. The scattering line shape at the O edge is described well by
O $p$ states hybrids with the Mn $e_{g}$ levels in the SMO and LMO
layers, shown with arrows in Fig. 2, followed at higher energy by
hybrids with La and Sr states. Two $e_{g,\downarrow}$ levels (3.5 eV
and 5 eV above $e_{g,\uparrow}$) are present in LMO and only one for
SMO (3.5 eV above $e_{g,\uparrow}$) (Ref.~\onlinecite{1995Saitoh})
because the electron in the $e_{g,\uparrow}$ level in LMO splits the
unoccupied $e_{g,\downarrow}$ levels by Coulomb interaction, even in
the absence of any Jahn-Teller distortion.

However, no variation across the FM transition is observed within
the error bars. To access T-dependent interface states (Sec. III A),
it is necessary to reach the $L=2$ reflection, for which the O $K$
edge energy is too low. In contrast, $L=2$ is accessible at the Mn
$L_{3,2}$ edges.

\begin{figure}
\centering\rotatebox{0}{\includegraphics[scale=0.5]{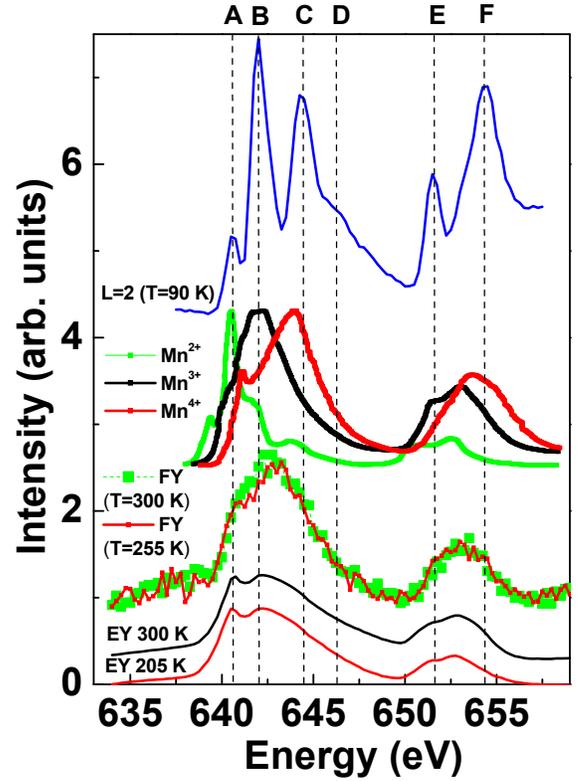}}
\caption{\label{fig:Figure1} (Color online) SL EY and FY (lower
curves) and RSXS (top) at $L=2$ at the Mn $L_{3,2}$ edges compared
to FY measurements for different Mn valences (middle, from
Ref.~\onlinecite{2004Morales}).}
\end{figure}

\subsubsection{Mn $L_{3,2}$ edge valences}

There is no discernible variation in the line shape at $L=1$ at the
Mn $L_{3,2}$ edges across the FM transition temperature (Fig. 4).
However, a pronounced variation is visible for $L=2$. From under the
relatively broad XAS at the Mn $L$ edge, T-dependent RSXS at $L=2$
selects those states that are sensitive to the temperature
variation. Specifically, an increased intensity of the $A$, $B$ and
$C$ peaks at lower T, and a decreased intensity of the $\alpha$ peak
is observed at the $L_{3}$ edge (Fig. 4). Parallel variations occur
at the $L_{2}$ edge for peaks $E$, $F$ and $\beta$. Similar results
were obtained for azimuthal angle $\phi=45^{\circ}$ (Fig. 5),
consistent with the azimuthal dependence of magnetic scattering. The
temperature dependence of height and width of peak $C$ are shown in
Fig. 6.

Coulomb and exchange interactions for ground and RSXS excited states
are different at the Mn $L_{3,2}$ edges. An analysis of the RSXS
line shapes based on ground state calculations, similar to that at
the O $K$ edge, cannot be made. However, both FY and RSXS
measurements probe excited states and FY measurements on bulk
samples for different Mn valences will be used to identify the
valence of scatterers for different resonances in the RSXS line
shapes. This approach is supported by the relatively small
difference in energy between RSXS resonances for scatterers of the
same valence and different specific scattering contrasts ($\rm
\lesssim 1~eV$ between magnetic and orbital scattering for bulk
measurements~\cite{2009Staub,2011Zhou}) compared to the $\sim
2\rm~eV$ separation in energy between the main FY features for
different valences (Fig. 3).

FY measurements in Sec. II B showed that the $\rm Mn^{2+}$ valence
is absent in the SL. The $A$ and $B$ resonances in the RSXS line
shape for $L=2$ correspond to the $\rm Mn^{3+}$ valence, while
resonance $C$ lines up at the energy of FY edge for a $\rm Mn^{4+}$
valence. More T-dependent measurements on SL with different
superperiods are needed before a quantitative discussion of peak
$\alpha$.

\section{Discussion}

The model of Sec. III A relates the variation with temperature of
the scattering contrast for $L=2$ and the absence of variation for
$L=1$ to changes in the form factor $\delta f_{i}$ of interface
layers and interface ferromagnetism.

The variation of $\delta f_{i}=f_{i,\rm FM}-f_{i,\rm PM}$ with the
transition from the PM to the FM state for specific interface
magnetic and orbital x-ray scattering models is discussed in Sec.
III B.

\subsection{Interface ferromagnetism}

The line shape and structure factor at the Mn edge at $L=2$ (Fig. 4)
are made of two parts. The T-independent structure factor $S$ is
given by the SL structure, to which a T-dependent contribution
$\delta S$ is added with the transition to the FM state [Fig. 6(a)].

The scattering intensity follows the evolution with temperature of
the FM moment [Fig. 6(a)]. The possibility that the transition to
the FM state gives exclusively non-magnetic scattering contrast is
not supported by an analysis of the line shapes (Sec. III B). The
T-dependent scattering $\delta S$ is, at least partially, magnetic
in origin and, for simplicity, we discuss only x-ray resonant
magnetic scattering in this section. Orbital contributions to
temperature variations in line shapes are addressed in Sec. III B.

The scattering intensity is $I\propto \big|S+\delta S\big|^{2}$. The
scattering form factors $(f)^{mn}$ and $(\delta f)^{mn}$ are
tensors, which are multiplied with the final ($\hat{\epsilon}_{f}$)
and initial ($\hat{\epsilon}_{i}$) light polarization vectors. This
gives an overall factor which, for charge ($f$) and magnetic
($\delta f$) scattering, is
\begin{eqnarray}
S ~~\propto ~~\hat{\epsilon}^{*}_{f,m} (f)^{mn}\hat{\epsilon}_{i,n}
~~\propto~~ (\hat{\epsilon}^{*}_{f} \hat{\epsilon}_{i})~f(\omega)
\\ \nonumber
\\
\delta S ~~\propto~~\hat{\epsilon}^{*}_{f,m} (\delta
f)^{mn}\hat{\epsilon}_{i,n} ~~\propto~~
i[(\hat{\epsilon}^{*}_{f}\times
\hat{\epsilon}_{i})\hat{z}_{l}]~\delta f(\omega)
\end{eqnarray}
\noindent where $\hat{z}_{l}$ is the direction of the local moment
at site $l$ (Refs.~\onlinecite{1990Kao,1995Tonnerre}). The sum over
the in-plane sites $l$ for each layer is proportional to the
magnetization $\overrightarrow{M}$ of the layer. $f(\omega)$ and
$\delta f(\omega)$ are scalar functions.

RSXS peaks that persist above the FM transition temperature are due
to scattering contrast defined by the SL structure, which is held
constant by the internal field between $\rm Sr^{2+}$ and $\rm
La^{3+}$ ions arranged in the SL layers. With the form factors shown
Fig. 1(a) for ``interface" ($f_{i}$), ``near-interface" ($f_{ni}$),
``middle SMO" ($f_{S}$) and ``middle LMO" ($f_{L}$) layers, and
neglecting inter-diffusion roughness and structural differences
between SMO/LMO and LMO/SMO interfaces, the T-independent structure
factors $S(Q)=\sum_{l} f_{l}e^{iQz_{l}}$ at $L=1, 2$ are
\begin{eqnarray}
S(L=1)= -f_{L}+f_{S}+f_{i}-f_{ni} \\
S(L=2)=f_{L}+f_{S}-f_{i}-f_{ni}
\end{eqnarray}
\noindent The origin has been chosen so that an arbitrary phase
factor between $S(L=1)$ and $S(L=2)$ is zero.

\begin{figure}
\centering\rotatebox{0}{\includegraphics[scale=0.55]{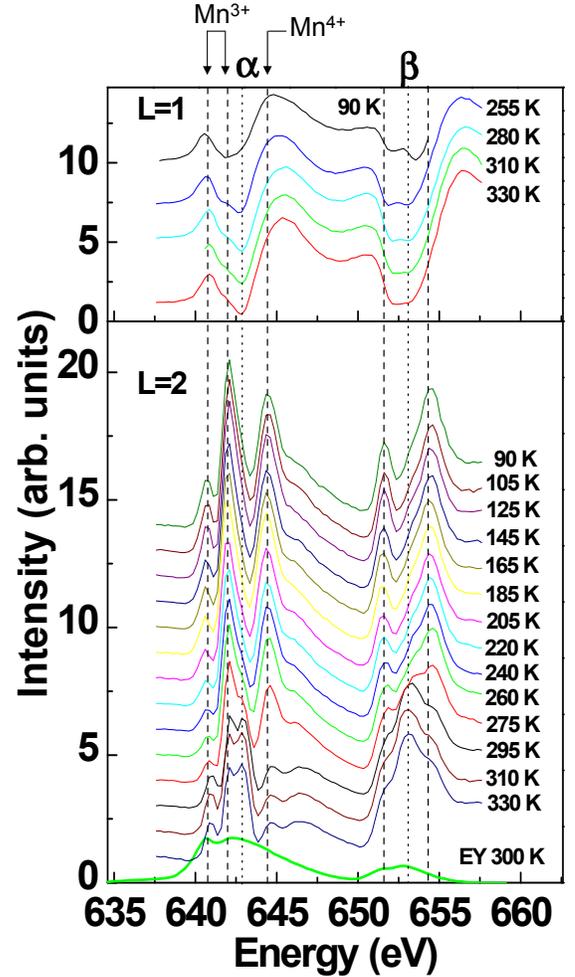}}
\caption{\label{fig:Figure1} (Color online) Temperature dependence
of RSXS at the Mn $L_{3,2}$ edges at $L=1$ and $L=2$ for azimuthal
angle $\phi=0^{\circ}$.}
\end{figure}

Higher momenta $L$ are not accessible at the Mn $L$ edges. However,
the absence of a variation with temperature in the scattering
intensity at $L=1$ [$\delta S(L=1)=0$] and the variation at $L=2$
[$\delta S(L=2)\neq 0$] strongly suggests that the unit cell of the
T-dependent contribution to the structure factor ($\delta S$) is
half the SL superperiod or 3 ML. The middle of the SMO and LMO
layers (separated by 3 ML) are the most dissimilar parts of the SL
structure, while the SMO/LMO and LMO/SMO interfaces (with two
interfaces every superperiod, also separated by 3 ML) are similar.
In the following, we consider a scattering component $\delta S$ that
develops at the SL interfaces.

These conditions are contrary to those expected for scattering from
crystal field effects or structural differences in a SL, from either
differences in the $c$-axis lattice constant or Jahn-Teller
distortions, which do not have a 3 ML unit cell. In addition, no
discernible variation was observed for scattering at the La
$M_{5,4}$ edges for $L=1$ or $L=2$ between 300 K and 225 K (data not
shown). This shows that the change in the line shape is due to a
variation with temperature in the resonant form factors of the SL
layers ($\delta f_{l}$), not of structural factors ($\delta z_{l}$).

In addition, these symmetry conditions on $\delta S$ set more
stringent constraints on the variation of $\delta f_{l}$ in the SL,
beyond the experimental observation $\delta S(L=1)=-\delta
f_{L}+\delta f_{S}+\delta f_{i}-\delta f_{ni}=0$. To obtain a 3 ML
unit cell, the variation in the middle of the LMO and SMO layers
must be equal, $\delta f_{L}=\delta f_{S}$. In addition, the
variation in the interface and near-interface layers must be equal,
$\delta f_{i}=\delta f_{ni}$ [Fig. 1(a)].

FM order in the $i$ and $ni$ layers is consistent with estimates of
the average Mn valence in a $\rm MnO_{2}$ plane based on the type
($L$ or $S$) of neighboring planes [Fig. 1(a)]. Specifically, a
comparison to magnetic orders of equivalent bulk LSMO doping shows
that $\rm Mn^{3+}$ and $\rm Mn^{3.5+}$ valences are near the FM dome
for bulk LSMO. The magnitude of the FM moment depends on the Mn
valence and implicitly on the SL interface roughness, with
structural imperfections in a $\rm (SMO)_{4.4}/(LMO)_{11.8}$ SL
correlating with the average interface FM moment.~\cite{2008May}
However, the FM moment distribution is more symmetrical in the
smaller superperiod $n=3$ SL (Ref.~\onlinecite{2008Bhattacharya}),
consistent with the symmetric FM moment distribution in Fig. 1(a).

The 3 ML unit cell of T-dependent scattering shows that there must
be two regions within a superperiod which are different from the FM
$i$ and $ni$ layers. With one the middle of the SMO layers
($f_{S}$), the other must be the middle of the LMO layers ($f_{L}$).
The x-ray scattering measurements imply that the magnetic scattering
in these two layers is the same $(\delta f_{S})_{\rm mag}=(\delta
f_{L})_{\rm mag}$. Therefore, these layers have either the same
magnetization $\overrightarrow{M}$ or no magnetization at low
temperatures. The different hole doping of these layers does not
support the possibility of an equal magnetization. The remaining
possibility is that, as the SL is cooled and becomes FM in zero
applied field, there is no magnetization in both these layers, or
$(\delta f_{S})_{\rm mag}=(\delta f_{L})_{\rm mag}=0$. Therefore,
for the $n=2$ SL, the variation in magnetic scattering $(\delta
f)_{\rm mag}$ and FM phase are localized at the SL interfaces.

A model of the magnetic state for the FM SL is shown in Fig. 1(a),
where $I$ and $II$ represent magnetic phases of the $f_{S}$ and
$f_{L}$ layers in no applied field (in contrast, the polarized
neutron reflectivity measurements in
Refs.~\onlinecite{2008May,2011Santos} were made in applied fields).
Since the average magnetization is zero for both $I$ and $II$
phases, the magnetic scattering has a 3 ML period. There are several
different possible $I$ and $II$ phases: a PM phase, an ordered AFM
phase (for instance, a C-type or a G-type), or an irregular phase
with canted moments~\cite{2011Santos} pointing in different
directions in the sample regions with slight variations in local
doping~\cite{1960deGennes}, even though the moments in the $f_{i}$
and $f_{ni}$ layers are always parallel.

\begin{figure}
\centering\rotatebox{0}{\includegraphics[scale=0.5]{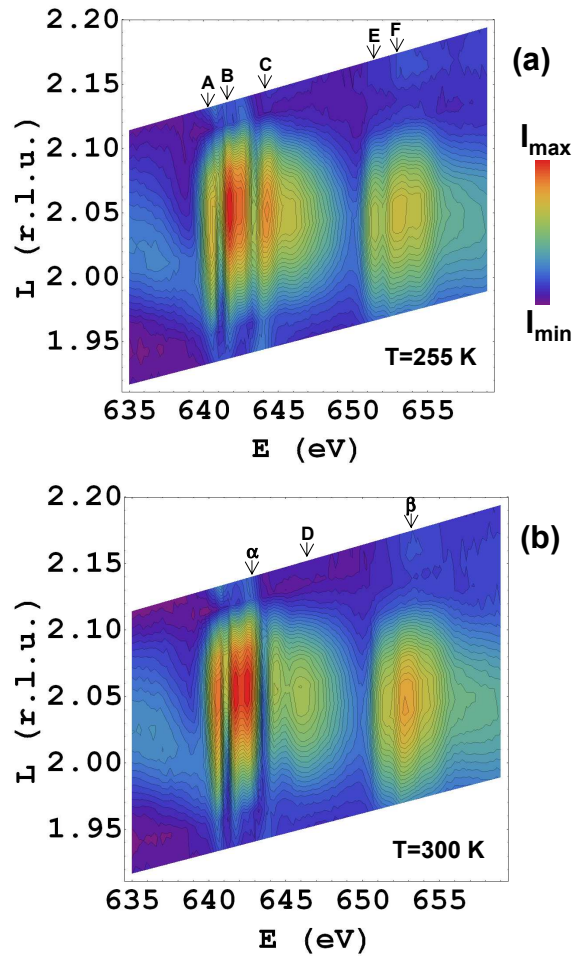}}
\caption{\label{fig:Figure1} (Color online) Two-dimensional
resonance profiles at the Mn $L_{3,2}$ edges at 255 K and 300 K for
azimuth $\phi=45^{\circ}$.}
\end{figure}

The valences of the $f_{L}$ and $f_{S}$ layers are close to $\rm
Mn^{3+}$ and $\rm Mn^{4+}$, which correspond to A-type and G-type
AFM magnetic orders in bulk LMO and SMO, with transition
temperatures of $\rm T_{\rm N, LMO}=135~K$
(Ref.~\onlinecite{1997Topfer}) and $\rm T_{\rm N, SMO} = 235~K$
(Ref.~\onlinecite{2001Chmaissem}), respectively. This suggests PM
$I$ or $II$ phases, at least for the higher temperature range, below
the SL FM transition temperature of $\rm 305~K$ [Fig. 6(a)].
However, the SL saturation FM moment of $\sim 2.5~\mu_{B}$ at $\rm
5~K$ (Ref.~\onlinecite{BNC}) gives the extent along the $c$-axis of
the FM region in high fields of $\gtrsim (2.5/3.22)\times 6~\rm
ML\sim 4.65~ML$ for each superperiod, where $\sim 3.22~\mu_{B}$ is
the maximum FM moment of the $x=0.33$ alloy~\cite{2008Bhattacharya}.
This value is too high for both $I$ and $II$ phases to remain PM at
the lowest temperatures. Therefore, at least one magnetic
transition, other than the FM transition, occurs in the SL.

The RSXS measurements are consistent with transitions in the $f_{S}$
and $f_{L}$ layers from a PM phase at higher temperatures to either
a G-type or C-type AFM phase (near the $\rm Mn^{4+}$ doping of bulk
LSMO) and to an irregular canted phase (near the the $\rm Mn^{3+}$
doping~\cite{1960deGennes}) at lower temperatures, respectively.
However, since there is no average layer magnetization in all these
cases, the RSXS intensity does not vary at these transitions, in
contrast to the FM transition in the $i$ and $ni$ layers.

With these constraints on $\delta f_{l}$, the change in the
structure factor $\delta S(Q)=\sum_{l} \delta f_{l}e^{iQz_{l}}$ at
the FM transition and $L=1, 2$ from Eqs. (3)-(4) is
\begin{eqnarray}
\delta S(L=1)=0\\
\delta S(L=2)=-2\delta f_{i}
\end{eqnarray}
\noindent Eq. (6) relates the changes with temperature in the line
shape at $L=2$ to variations of form factor of interface $\delta
f_{i}$ and near-interface $\delta f_{ni}=\delta f_{i}$ layers in the
SL.

The $\delta S(L=2)$ reflection is allowed in this $n=2$ SL for all
Mn sites in the FM layers. However, similar to AFM orders in bulk
LSMO, scattering from a $\rm Mn^{4+}$ valence was not observed at
$L=3$ for a $n=4$ SL (Ref.~\onlinecite{2007S}) [it was observed in a
$n=3$ SL at $L=3$ (data not shown)]. The symmetry that very
effectively forbids reflections from the $\rm Mn^{4+}$ ions at $L=3$
for the $n=4$ SL is not known and surprising, given inherent small
imperfections of a SL structure. More measurements are needed for
different SL to answer this question.

\subsection{Interface x-ray scattering}

We now discuss the temperature variation of the RSXS line shape at
$L=2$ and Mn $L_{3,2}$ edges.

The width of resonance $C$, corresponding to the $\rm Mn^{4+}$
valence and to interface and near-interface layers, has an sharp
increase at the FM transition temperature [Fig. 6(b)]. The increase
in the scattering intensity in the FM state is also taking place
$\sim 0.2~\rm eV$ below the charge order resonance that corresponds
to the $\rm Mn^{4+}$ valence in the PM state (at $\rm 644.65~eV$ in
Fig. 4).

In general, the line shape of resonant magnetic scattering is
related to variations in the occupation of orbitals induced by a
magnetic field~\cite{1988Hannon} near the FY edges for Mn ions of
different ($\rm Mn^{3+}$ and $\rm Mn^{4+}$) valences. However, the
magnetic scattering is slightly shifted to lower energies compared
to orbital scattering for AFM bulk orders.~\cite{2009Staub,2011Zhou}
We cannot resolve two peaks at $C$ in the SL line shapes at low T,
but this suggests that, with the increase of the FM moment at lower
T, a T-dependent magnetic scattering contribution is added $\sim
0.2~\rm eV$ below the charge scattering resonance. This addition to
$f_{i}$ of a temperature dependent $(\delta f_{i})_{\rm mag}$
explains the observed variation in line shape at $L=2$. The charge
scattering resonance might also increase at lower T, concomitantly
with magnetic scattering and variations in orbital scattering with T
are discussed briefly at the end of this Section.

A more gradual increase in width is observed at lower T [Fig. 6(b)].
For x-ray scattering in the FM phase, it is necessary to consider a
double-exchange two-site orbital, which suggests that this width
increase is related to the T dependence of the double-exchange
frequency $t_{ij}$ between the two Mn sites. Both resonant magnetic
and orbital scattering are ultimately scattering off orbitals, and
the consideration of two-site orbitals in the FM state applies to
both cases.

\begin{figure}
\centering\rotatebox{0}{\includegraphics[scale=0.45]{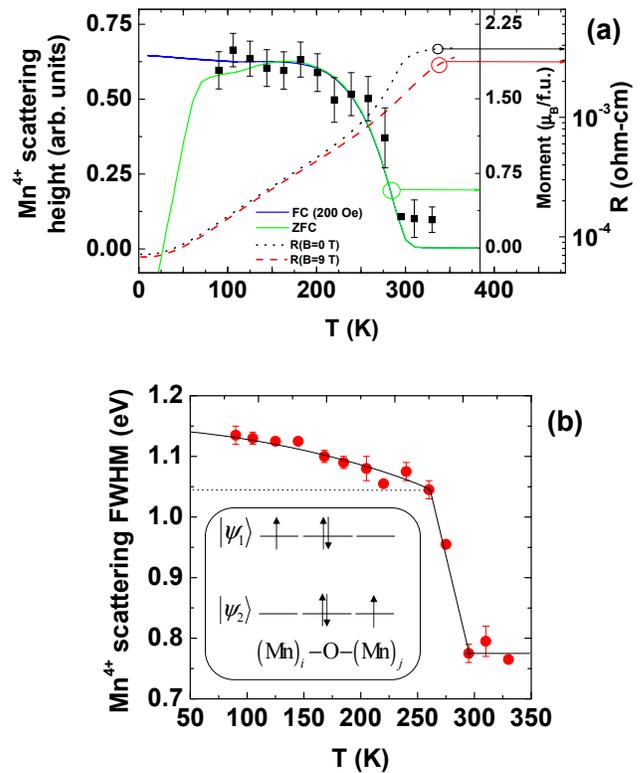}}
\caption{\label{fig:Figure1} (Color online) (a) Temperature
dependence of resonance $C$ height for $\phi=0^{\circ}$ compared to
the SL FM moment measured with SQUID for ZFC and in-plane FC=200 Oe
(Ref.~\onlinecite{BNC}). The SL has a $\rm 305\pm 5~K$ FM transition
temperature, which is lower than the $\rm \sim 355~K$ transition
temperature of the $x=0.33$ LSMO alloy
(Ref.~\onlinecite{2008Bhattacharya}). Hysteresis loops show that the
easy axis is in-plane (data not shown). SL resistance becomes
metallic-like at low T (Ref.~\onlinecite{BNC}). (b) Temperature
dependence of resonance $C$ width. The line is a guide to the eye.
Inset shows a sketch of the double-exchange configurations
$|\psi_{1}\rangle$ and $|\psi_{2}\rangle$ for Mn sites $i$ and $j$
(Ref.~\onlinecite{1951Zener}).}
\end{figure}

The double-exchange process involves two coordinated jumps from the
Mn to the O atoms [Fig. 6(b), inset]. It is useful to consider the
simpler process of one jump first, which is sometimes included in
XAS calculations of complex oxides. In this case, inter-site charge
transfer between $d$-states of a transition metal and a neighboring
(ligand, $L$) O ion~\cite{2005Groot} and consideration of multiple
configurations (for instance, $d^{8}$ and $d^{9}\underline{L}$ for
$\rm Cu^{1+}$ and $\rm Cu^{2+}$ valences) change the scattering form
factor $f$ at the transition metal edge. In particular, satellite
peaks develop in XAS (and, implicitly, in RSXS) at additional Cu
valences.~\cite{1998Hu}

In the double-exchange process, specific to FM complex oxides,
charge transfer takes place between transition metal sites, beyond
the neighboring O atoms. Specifically, the $|\psi_{1}\rangle$ and
$|\psi_{2}\rangle$ configurations are coupled in a two-site ground
state wave function [Fig. 6(b), inset], which for this FM manganite
is
\begin{eqnarray}
|\psi_{{\pm}}\rangle=\frac{1}{\sqrt{2}}(|\psi_{1}\rangle \pm
|\psi_{2}\rangle)\\ \nonumber |\psi_{1}\rangle=|\rm
{Mn^{3+},O^{2-},Mn^{4+}}\rangle\\ \nonumber |\psi_{2}\rangle=|\rm
{Mn^{4+},O^{2-},Mn^{3+}}\rangle
\end{eqnarray}
\noindent with Mn valences in FM layers in a superposition of $\rm
Mn^{3+}$ and $\rm Mn^{4+}$. In the ground state (without an x-ray
photon absorbed), the charge transfer splits the two levels
$|\psi_{\pm}\rangle$ by the exchange energy $2t_{ij}$
(Ref.~\onlinecite{1951Zener}), where the double-exchange hopping
between sites $i$ and $j$ is $t_{ij}=t_{\rm DE} \rm
cos[(\theta_{i}-\theta_{j})/2]$, with $t_{\rm DE}$ a constant and
$\theta_{i,j}$ the $t_{2g,\uparrow}$ spins orientations on the two
sites (Ref.~\onlinecite{1955Anderson}).

To account for the double-exchange process in x-ray scattering, the
orbitals $|\psi_{1,2}\rangle$ are replaced with the two-site
orbitals $|\psi_{\pm}\rangle$. Similar to the case of ligand holes
on oxygen atoms, the charge transfer between Mn sites beyond the
neighboring O atoms changes the scattering factor $f$ at satellite
peaks in RSXS, which correspond to the $\rm Mn^{3+}$ and $\rm
Mn^{4+}$ valences.

In addition, the splitting by $2t_{ij}$ of the $|\psi_{\pm}\rangle$
states results, for our relatively large energy resolution and core
hole width, in an increase of the measured width by $2t_{ij}$. More
precisely, the the bandwidth of $e_{g}$ electrons depends on the
hopping frequency between the $i$ and $j$ sites as~\cite{1997Topfer}
\begin{eqnarray}
W\propto \rm  cos[(\theta_{i}-\theta_{j})/2] cos \phi \propto
\it{t_{ij}} \rm cos \phi
\end{eqnarray}
\noindent where $(\pi - \phi)$ is the angle between the Mn-O-Mn
bonds. A width increase at lower T in the ground state is
transferred to a increase of the RSXS line width.

The hopping frequency $t_{ij}$ increases with increased FM order of
spins $\theta_{\rm i,j}$ at lower T, and broadens the scattering
form factors $f$ and the line width. In this model, the XAS and
therefore, the RSXS peaks, should become broader at lower
temperatures. The width increase at lower temperatures of peak $C$
[Fig. 6(b)] is consistent with this model and $t_{\rm band}\sim
0.2-0.5~\rm eV$ for each of the $e_{g}$ states, $2t_{\rm DE}\sim
2T_{\rm Curie} \sim 0.05~\rm eV$ and contributions from experimental
resolution ($\rm 0.34~eV$ at the Mn $L$ edge) and core-hole width
($w_{\rm FWHM} \rm \sim 0.3-0.5~eV$, Ref.~\onlinecite{2004Thomas}).

In addition to the double-exchange processes in the FM state,
lattice distortions are also relevant to the CMR
transition~\cite{1995Hwang,1996Millis}. In bulk manganites, they may
depend on T, changing the bond alignment $\phi$ and bandwidth $W$
[Eq. (8)]. However, the average angle between the Mn-O-Mn bonds for
SL samples is fixed by the substrate.

Orbital scattering at the Mn $L$ edges has a comparable amplitude to
magnetic scattering for bulk AFM orders.~\cite{2009Staub,2011Zhou}
It can come from occupation contrast or polarization contrast from
different atomic orbital orientations in the anomalous scattering
tensor. The analogous occupation contrast in SL FM is a T-dependent
charge transfer across SL interfaces (which includes the electronic
reconstruction of Ref.~\onlinecite{2007S}), in addition to the
T-independent part defined by the SL structure. The T-dependent
polarization contrast in the SL may also be substantial; for
instance, on closely related SL (Ref.~\onlinecite{2009Aruta}),
in-plane $e_{g}(x^{2}-y^{2})$ occupation and FM near LMO interfaces
and out-of-plane $e_{g}(3z^{2}-r^{2})$ orbital occupation and AFM in
the middle of LMO layers was inferred from XMLD and XMCD
measurements. Polarization-resolved scattering measurements in a
magnetic field with $\pi$ and $\sigma$ incident light and scattered
beam polarization analysis are necessary to separate different
magnetic and orbital contributions to scattering at the Mn $L_{3,2}$
edges.

We discuss the O $K$ edge briefly. Oxygen doping is consistent with
our observations (Fig. 2), other measurements~\cite{1997Ju} and
certain models~\cite{2010GF}. The interface FM state of this $n=2$
SL is the metallic state observed in a $n=4$ SL
(Ref.~\onlinecite{2007S}). The reflection at $L=1$ (Fig. 2) is not
sensitive to T-dependent scattering because, as for the Mn $L$
edges, $\delta S(L=1)=0$. The intensity of the $L=3$ reflection in
Ref.~\onlinecite{2007S} corresponds to the $L=2$ reflection for this
SL. In particular, to determine whether T-dependent scattering
occurs at the O $K$ edge in this SL, it would be necessary to
measure at $L=2$, a momentum which is not accessible at the O $K$
edge.

\section{Conclusion}

X-ray absorption measurements at the O $K$ edge in a $\rm
SrMnO_{3}/LaMnO_{3}$ superlattice showed a shoulder, corresponding
to holes doped on oxygen sites. The shoulder is aligned with the
main resonant peak of soft x-ray scattering from the spatial
modulation in the density of doped holes.

A large variation in the Mn $L_{3,2}$ line shapes at $L=2$, but not
at $L=1$, was observed across the FM transition, pointing to
scattering from ferromagnetic interfaces.

Comparison to fluorescence yield edge energies for different Mn
valences showed the presence of scattering contrast at both $\rm
Mn^{3+}$ and $\rm Mn^{4+}$ valences. An x-ray scattering model,
which includes double-exchange orbitals in the FM state, explains
the observed line broadening at lower temperatures.

Measurements on bulk 113 (Ref.~\onlinecite{2004Thomas}), 214
(Ref.~\onlinecite{2003Wilkins-b,2004Dhesi,2005Wilkins}) and 327
(Ref.~\onlinecite{2006Wilkins}) manganites in the AFM state observe
two main resonances only at the Mn $L_{3}$ edge, at $A$ and $B$.
Scattering from $\rm Mn^{4+}$ ions (corresponding to resonance $C$)
has the symmetry of a forbidden reflection. Specifically, it has a
spatial periodicity of 2 u.c. and is not allowed at the bulk
in-plane orbital order reflection wave vector of 4 u.c. along the
tetragonal axes (Ref.~\onlinecite{2003Wilkins-b,2004Thomas}). Having
to rely on measurements of the $\rm Mn^{3+}$ resonances only,
different methods to determine the charge disproportionation for
bulk AFM orders are controversial, with both small and large charge
disproportionation obtained. Our RSXS line shapes, for a SL
structure with a large intrinsic charge disproportionation, add an
experimental constraint on these competing models.

The development of the SL FM order was accessed with x-ray resonant
magnetic scattering and no applied magnetic fields. An open question
is the trace [FC or ZFC in Fig. 6(a)] that the height of resonance
$C$ would follow on further cooling.

We would like to contrast our measurements to polarized neutron
reflectivity (PNR) data on SMO/LMO superlattices
(Refs.~\onlinecite{2008May,2011Santos}), where a magnetic modulation
was measured with a period equal to the SL superperiod
(magnetization strongly suppressed in SMO, high in LMO). In
contrast, the RSXS measurements presented here show an ordering of
magnetic moments with a period equal to half the SL superperiod.
Several factors may be at the origin of this difference. First, the
experimental conditions of the PNR and RSXS measurements were
different. Specifically, PNR measurements were made in relatively
high fields ($\rm 0.55~T$ and $\rm 0.82~T$ in
Ref.~\onlinecite{2008May} and~\onlinecite{2011Santos},
respectively), while the RSXS measurements were made with no applied
fields. Second, the samples measured in this work have a lower SL
superperiod ($n=2$) compared to the samples of PNR measurements
($n=3$ and $n=5$). Thus, a complete mapping of the magnetic
structure of SMO/LMO superlattices as a function of deposition
sequence, magnetic field and temperature requires more measurements.

\section{Acknowledgments}

This work was supported by the Department of Energy Office of Basic
Energy Science: RSXS measurements by grant DE-FG02-06ER46285, NSLS
facilities by DE-AC02-98CH10886, and MRL facilities by
DE-FG02-07ER46453 and DE-FG02-07ER46471. Work at Argonne National
Laboratory, including use of facilities at the Center for Nanoscale
Materials, was supported by the U.S. Department of Energy, Office of
Basic Energy Sciences under contract No. DE-AC02-06CH11357.


\begin{thebibliography}{}

\bibitem{1955Wollan}
E.O. Wollan and W.C. Koehler, Phys. Rev. \textbf{100}, 545 (1955).

\bibitem{1955Goodenough}
J.B. Goodenough, Phys. Rev. \textbf{100}, 564 (1955).

\bibitem{1960deGennes}
P.-G. de Gennes, Phys. Rev. \textbf{118}, 141 (1960).

\bibitem{1951Zener}
C. Zener, Phys. Rev. \textbf{82}, 403 (1951).

\bibitem{1955Anderson}
P.W. Anderson and H. Hasegawa, Phys. Rev. \textbf{100}, 675 (1955).

\bibitem{1950Anderson}
P.W. Anderson, Phys. Rev. \textbf{79}, 350 (1950).

\bibitem{1994Jin}
S. Jin, T.H. Tiefel, M. McCormack, R.A. Fastnacht, R. Ramesh, and
L.H. Chen, Science \textbf{264}, 413 (1994).

\bibitem{2004Thomas}
K.J. Thomas \emph{et al.}, Phys. Rev. Lett. \textbf{92}, 237204
(2004).

\bibitem{2003Wilkins-b}
S.B. Wilkins \emph{et al.}, Phys. Rev. Lett. \textbf{91}, 167205
(2003).

\bibitem{2004Dhesi}
S.S. Dhesi \emph{et al.}, Phys. Rev. Lett. \textbf{92}, 056403
(2004).

\bibitem{2005Wilkins}
S.B. Wilkins \emph{et al.}, Phys. Rev. B \textbf{71}, 245102 (2005).

\bibitem{2006Wilkins}
S.B. Wilkins \emph{et al.}, J. Phys.: Condens. Matter \textbf{18},
L323 (2006).

\bibitem{2009Staub}
U. Staub \emph{et al.}, Phys. Rev. B \textbf{79}, 224419 (2009).

\bibitem{2010GF}
M. Garcia-Fernandez \emph{et al.}, Phys. Rev. B \textbf{82}, 235108
(2010).

\bibitem{2011Zhou}
S.Y. Zhou \emph{et al.}, Phys. Rev. Lett. \textbf{106}, 186404
(2011).

\bibitem{2011Ehrke}
H. Ehrke \emph{et al.}, Phys. Rev. Lett. \textbf{106}, 217401
(2011).

\bibitem{2005Stojic}
N. Stojic, N. Binggeli, and M. Altarelli, Phys. Rev. B \textbf{72},
104108 (2005).

\bibitem{2000Castleton}
C.W.M. Castleton and M. Altarelli, Phys. Rev. B \textbf{62}, 1033
(2000).

\bibitem{2010OBthesis}
O. Bunau, PhD thesis, University of Grenoble (2010).

\bibitem{2007S} S. Smadici \emph{et al.},
Phys. Rev. Lett. \textbf{99}, 196404 (2007).

\bibitem{2001Chmaissem}
O. Chmaissem \emph{et al.}, Phys. Rev. B \textbf{64}, 134412 (2001).

\bibitem{1998Rodriguez-Carvajal}
J. Rodriguez-Carvajal \emph{et al.}, Phys. Rev. B \textbf{57}, R3189
(1998).

\bibitem{2011S}
S. Smadici, J.C.T. Lee, J. Morales, G. Logvenov, O. Pelleg, I.
Bozovic, Y. Zhu and P. Abbamonte, Phys. Rev. B \textbf{84}, 155411
(2011).

\bibitem{1997Ju}
H.L. Ju, H-C. Sohn, and K.M. Krishnan, Phys. Rev. Lett. \textbf{79},
3230 (1997).

\bibitem{1991Cramer}
S. P. Cramer \emph{et al.}, J. Am. Chem. Soc. \textbf{113}, 7937
(1991).

\bibitem{2004Morales}
F. Morales \emph{et al.}, J. Phys. Chem. B \textbf{108}, 16201
(2004).

\bibitem{2009Lee}
J. Lee \emph{et al.}, Phys. Rev. B \textbf{80}, 205112 (2009).

\bibitem{2005Abbamonte}
P. Abbamonte, A. Rusydi, S. Smadici, G.D. Gu, G.A. Sawatzky, and
D.L. Feng, Nature Physics \textbf{1}, 155-158 (2005).

\bibitem{2011Lee}
J.-S. Lee, C.-C. Kao, C.S. Nelson, H. Jang, K.-T. Ko, S.B. Kim, Y.J.
Choi, S.-W. Cheong, S. Smadici, P. Abbamonte, and J.-H. Park, Phys.
Rev. Lett. \textbf{107}, 037206 (2011).

\bibitem{2009S}
S. Smadici, J.C.T. Lee, S. Wang, P. Abbamonte, G. Logvenov, A.
Gozar, C. Deville Cavellin, and I. Bozovic, Phys. Rev. Lett.
\textbf{102}, 107004 (2009).

\bibitem{2012S}
S. Smadici, J.C.T. Lee, A. Rusydi, G. Logvenov, I. Bozovic, and P.
Abbamonte, Phys. Rev. B \textbf{85}, 094519 (2012).

\bibitem{1990Kao}
C.C. Kao \emph{et al.}, Phys. Rev. Lett. \textbf{65}, 373 (1990).

\bibitem{1994Kao}
C.C. Kao \emph{et al.}, Phys. Rev. B \textbf{50}, 9599 (1994).

\bibitem{1995Tonnerre}
J.M. Tonnerre \emph{et al.}, Phys. Rev. Lett. \textbf{75}, 740
(1995).

\bibitem{1992Abbate}
M. Abbate \emph{et al.}, Phys. Rev. B \textbf{46}, 4511 (1992).

\bibitem{1995Saitoh}
T. Saitoh \emph{et al.}, Phys. Rev. B \textbf{51}, 13942 (1995).

\bibitem{1997Topfer}
J. Topfer \emph{et al.}, J. of Solid State Chemistry \textbf{130},
117 (1997).

\bibitem{2008May}
S. May \emph{et al.}, Phys. Rev. B \textbf{77}, 174409 (2008).

\bibitem{2008Bhattacharya}
A. Bhattacharya \emph{et al.}, Phys. Rev. Lett. \textbf{100}, 257203
(2008).

\bibitem{BNC}
B.B. Nelson-Cheeseman \emph{et al.} (unpublished).

\bibitem{2011Santos}
T. Santos \emph{et al.}, Phys. Rev. Lett. \textbf{107}, 167202
(2011).

\bibitem{1988Hannon}
J.P. Hannon, G.T. Trammell, M. Blume, and D. Gibbs, Phys. Rev. Lett.
\textbf{61}, 1245 (1988).

\bibitem{2005Groot}
F. de Groot, Coordination Chemistry Reviews \textbf{249}, 31 (2005).

\bibitem{1998Hu}
Z. Hu \emph{et al.}, Chemical Physics \textbf{232}, 63 (1998).

\bibitem{1995Hwang}
H.Y. Hwang, S.W. Cheong, P.G. Radaelli, M. Marezio, and B. Batlogg,
Phys. Rev. Lett. \textbf{75}, 914 (1995).

\bibitem{1996Millis}
A.J. Millis, R. Mueller, and B.I. Shraiman, Phys. Rev. B
\textbf{54}, 5405 (1996).

\bibitem{2009Aruta}
C. Aruta \emph{et al.}, Phys. Rev. B \textbf{80}, 140405(R) (2009).

\end{thebibliography}
\end{document}